\documentclass[a4paper]{article}
\usepackage{Odyssey2020}
\usepackage{epsfig,amssymb,amsmath}
\usepackage{algorithm}
\usepackage{algorithmicx}
\usepackage{algpseudocode}
\usepackage{amsmath}
\usepackage{enumerate}
\usepackage{booktabs}
\usepackage{multirow}
\usepackage[export]{adjustbox}
\newcommand{\rom}[1]{\uppercase\expandafter{\romannumeral #1\relax}}
\usepackage{setspace}
\ninept

\title{DIHARD \rom{2} is Still Hard: Experimental Results and Discussions\\from the DKU-LENOVO Team}

\name{
 Qingjian Lin$^{1,2}$, 
 Weicheng Cai$^{1,2}$, 
 Lin Yang$^4$,
 Junjie Wang$^4$,
 Jun Zhang$^2$,
 Ming Li$^{1,3}$
 }

\address{
 $^1$Data Science Research Center, Duke Kunshan University, Kunshan, China \\
 $^2$School of Electronics and Information Technology, Sun Yat-sen University, Guangzhou, China \\
 $^3$School of Computer Science, Wuhan University, Wuhan, China \\
 $^4$AI Lab of Lenovo Research, Beijing, China \\
 {\small \tt ming.li369@dukekunshan.edu.cn} 
}
\begin{document}
\maketitle

\begin{abstract}
In this paper, we present the submitted system for the second DIHARD Speech Diarization Challenge from the DKU-LENOVO team. Our diarization system includes multiple modules, namely voice activity detection (VAD), segmentation, speaker embedding extraction, similarity scoring, clustering, resegmentation and overlap detection. For each module, we explore different techniques to enhance performance. Our final submission employs the ResNet-LSTM based VAD, the Deep ResNet based speaker embedding, the LSTM based similarity scoring and spectral clustering. Variational Bayes (VB) diarization is applied in the resegmentation stage and overlap detection also brings slight improvement. Our proposed system achieves 18.84\% DER in Track1 and 27.90\% DER in Track2. Although our systems have reduced the DERs by 27.5\% and 31.7\% relatively against the official baselines, we believe that the diarization task is still challenging. 

\noindent\textbf{Index Terms}: DIHARD, VAD, speaker embedding, similarity scoring, clustering, resegmentation, overlap detection
\end{abstract}

\section{Introduction}
Speaker diarization is the task of determining ``who spoke when'' in an audio file that usually contains an unknown number of speakers with variable speech duration~\cite{tranter2006overview,anguera2012speaker}. It has a wide range of applications such as telephone calls, meeting recordings and broadcast interviews. Diarization can also serve as the frontend of automatic speech recognition (ASR) to improve the transcription performance in multi-speaker conversations. 

In general, a diarization system partitions multi-speaker audios into short segments and clusters them according to speaker identities. It consists of following modules (shown in Figure~\ref{fig:framework}):
\begin{figure}[!htb]
 \includegraphics[width=\linewidth]{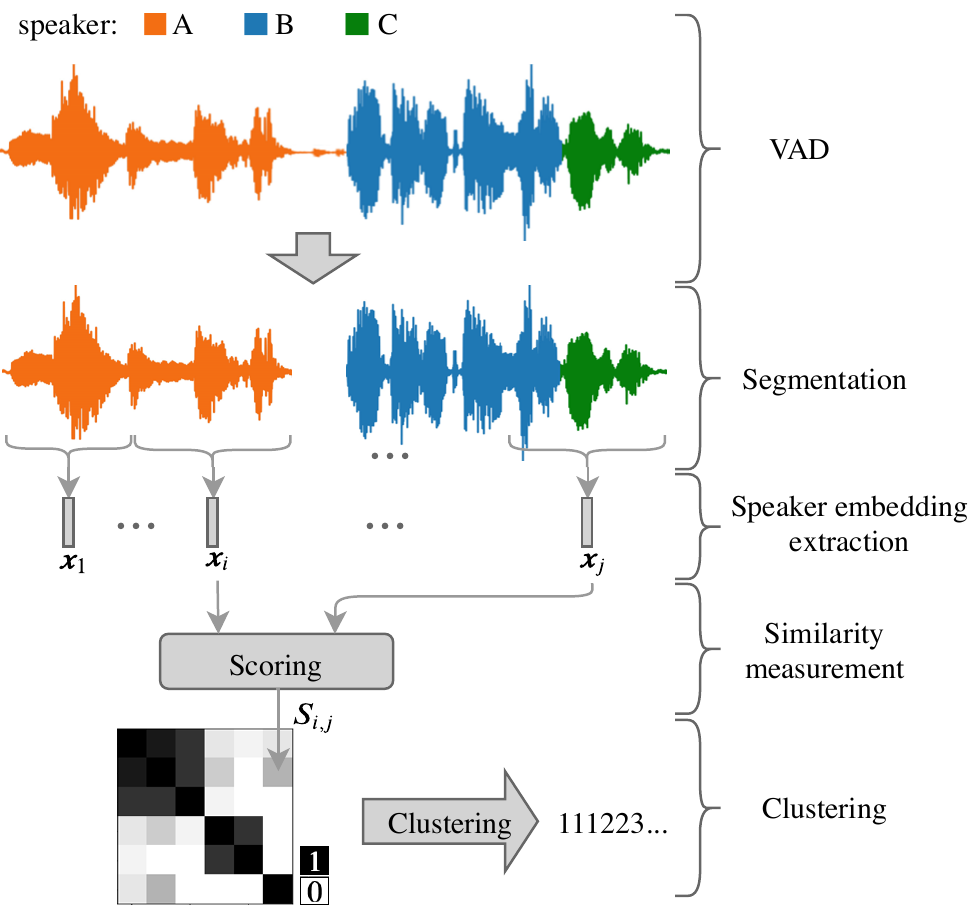}
 \caption{A standard speaker diarization framework with multiple modules.}
 \label{fig:framework}
 \vspace{-3mm}
\end{figure}

\begin{table*}[]
 \vspace{-4mm}
 \centering
 \caption{Metadata analysis on the DIHARD \rom{2} dev set.}
 \label{tab:data}
 \vspace{2mm}
 \begin{tabular}{cccccc}
 \toprule
 domains & n\_audios & n\_speakers & average duration & speech percentage(\%) & overlapped error(\%) \\
 \midrule
 restaurant & 12 & 5$\sim$8 & 10min 6s & 88.01 & 25.20 \\
 meeting & 14 & 3$\sim$10 & 10min 27s & 93.89 & 22.42 \\
 webvideo & 32 & 1$\sim$9 & 3min 31s & 75.12 & 21.70 \\
 child & 23 & 2$\sim$5 & 9min 51s & 59.78 & 11.73 \\
 socio\_field & 12 & 2$\sim$6 & 10min 2s & 72.62 & 7.53 \\
 socio\_lab & 16 & 2 & 5min 59s & 74.38 & 4.78 \\
 clinical & 24 & 2$\sim$3 & 3min 26s & 60.94 & 3.75 \\
 maptask & 23 & 2 & 6min 33s & 68.15 & 2.92 \\
 court & 12 & 5$\sim$10 & 10min 24s & 84.09 & 1.90 \\
 broadcast\_interview & 12 & 3$\sim$5 & 10min 17s & 78.70 & 1.18 \\
 audiobooks & 12 & 1 & 10min 3s & 79.37 & 0.00 \\
 \midrule
 ALL & 192 & 1$\sim$10 & 7min 25s & 76.07 & 10.76 \\
 \bottomrule
 \end{tabular}
 \vspace{-2mm}
\end{table*}

\vspace{-3mm}
\begin{itemize}
 \item \textbf{Voice activity detection (VAD)}: VAD detects speech in the audio signals and removes the non-speech regions to reduce computation. Typical VAD systems include generative models like Hidden Markov Model (HMM)~\cite{wooters2004towards}, and discriminative models like linear discriminant analysis (LDA)~\cite{rentzeperis20062006}, support vector machine (SVM)~\cite{temko2007enhanced} and deep neural network (DNN) methods~\cite{zazo2016feature,eyben2013real,chang2018temporal}.
 
 \item \textbf{Segmentation}: The segmentation step splits speech into multiple speaker-homogeneous segments. In general, a speaker changepoint detector (SCD)~\cite{hruz2017convolutional, yin2017speaker} is employed to search speaker changepoints and split speech by the changepoints, but uniform segmentation with overlap~\cite{sell2014speaker} also works fine. 

 \item \textbf{Speaker embedding extraction}: After segmentation, short segments are mappedd into the speaker subspace and generate fixed-dimensional speaker embeddings such as i-vector~\cite{shum2010unsupervised}, x-vector~\cite{snyder2018x} and Deep ResNet based speaker embeddings~\cite{cai2018exploring,cai2018analysis}.

 \item \textbf{Similarity measurement}: Similarity scores between any two speaker embeddings in the same audio are computed and later used in the clustering step. Popular techniques includes cosine similarity, probabilistic linear discriminant analysis (PLDA)~\cite{prince2007probabilistic,kenny2013plda} and DNN based measurement~\cite{lin2019lstm}.

 \item \textbf{Clustering}: Clustering algorithms like K-means~\cite{wang2018speaker}, agglomerative hierarchical clustering (AHC)~\cite{sell2014speaker, sell2018diarization} and spectral clustering~\cite{lin2019lstm, wang2018speaker} assign segments with high similarity scores to the same cluster. Note that the similarity measurement and clustering steps can be merged by a single online clustering backend, according to~\cite{zhang2019fully,li2019discriminative}.
\end{itemize}
The aforementioned modules construct the general framework of speaker diarization. Additional modules like resegmentation~\cite{diez2018speaker,sell2015diarization} and overlap detection~\cite{novotny2018but} are not essential but help further improve the system performance.

Researches have achieved state-of-the-art performance in some eval datasets like CallHome~\cite{wang2018speaker,zhang2019fully}. However, the 2017 JSALT Summer Workshop at CMU found it hard to migrate the success to more challenging corpora including web videos, speech in the wild, child language recordings, etc~\cite{ryanta2018enhancement}. To raise researchers' attention, the DIHARD competition is therefore proposed as a new annual challenge focusing on the ``hard'' scenarios. The second DIHARD speech diarization challenge~\cite{ryant2019second,berg2016,ryant2019} contains four tracks. Track1 and Track2 share the same single-channel audios while Track3 and Track4 use the same multiple-channel audios. An oracle voice activity detector is provided in Track1 and Track3. As for the rest tracks, participants are required to distinguish between speech and nonspeech by themselves. Since we only participate in Track1 and Track2, experiments and discussions are limited in these two tracks.

The rest of this paper is organized as follows. Section 2 carries out metadata analysis on the DIHARD \rom{2} development (dev) set. Section 3 describes models and algorithms we used in the competition. Experimental results and discussions are presented in Section 4, while conclusions are drawn in Section 5. 

\section{Metadata analysis}
In this section, we carry out metadata analysis on the DIHARD \rom{2} dev set to illustrate how ``hard'' the competition is. Several indicators are important: duration of audiosvox, the number of speakers, speech percentage and ovelapped error. It is discussed in~\cite{lin2019lstm} that performance of diarization systems degrades as audio duration increases. Besides, with more speakers involved in conversations, it becomes harder for systems to make correct predictions about each speaker. Speech percentage reflects proportion of speech in audios and helps us choose suitable datasets for VAD training. Overlapped error determines the minimum diarization error rate (DER) a diarization system is possible to achieve without handling overlapped speech~\footnote{The DIHARD competition takes the strictest definition of DER: no colloar is tolerated around speech boundaries and overlapped speech accounts for DER.}. Assume that $D$ and $N$ denote duration and the number of speakers respectively, and $R_i$ denotes speech regions of specific speaker $i$. Then speech percentage $R_\%$ can be expressed as follows:

\begin{equation}
 R_\% = \frac{dur(R_1\cup R_2\cup \cdots \cup R_N)}{D}.
\end{equation}
The $dur(\cdot)$ function gets duration of input regions. Besides, the overlapped error $O_\%$ is calculated by:
\begin{equation}
 \begin{aligned}
 O_\% = \frac{\sum_{n=1}^Ndur(R_i) - dur(R_1\cup R_2\cup \cdots \cup R_N)}{\sum_{n=1}^Ndur(R_i)}.
 \end{aligned}
\end{equation}

Results are shown in Table~\ref{tab:data}. In comparison with existing datasets, the DIHARD \rom{2} dev set is drawn from 11 different domains. In some challenging domains like $restaurant$, conversations arise under the background of strong noise, making it difficult for systems to detect speakers. The number of involved speakers ranges from 1 to 10, and most of the domains contain conversations with more than two speakers except $audiobooks$, $maptask$ and $socio\_lab$. The average audio duration is 7min~25s, and three quarters is speech. The overall overlapped error is 10.76\%, but in specific domains such as $meeting$, $restaurant$ and $webvideo$, the error is much higher. It indicates that audios drawn from these domains are much harder to cope with, even in the DIHARD competition. In summary, the competition is ``hard'' mainly because:
\begin{enumerate}[1)]
 \item audios are drawn from a diverse set of challenging domains.
 \item the number of speakers varies in a large range. 
 \item high overlapped error accounts for DER.
\end{enumerate}

\begin{figure*}[t]
 \centering
 \includegraphics[width=0.9\linewidth]{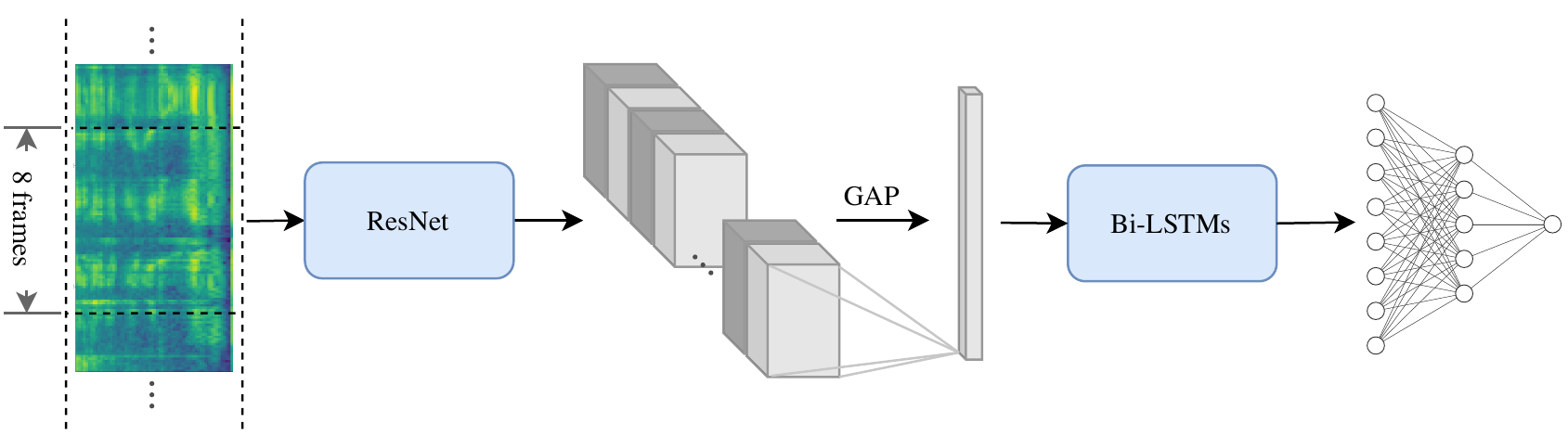}
 \caption{The structure of ResNet-LSTM VAD.}
 \label{fig:vad}
\end{figure*}

\section{System description}
\subsection{Data}
\label{sec:data}
Multiple datasets are employed in our experiments. Voxceleb1\&2~\cite{nagrani2017voxceleb} contain 16k sampled short utterances with single speakers, suitable for speaker embedding training. Multi-speaker audios are drawn from databases in the meeting and telephone domains. The 16k sampled meeting data consists of AMI~\cite{carletta2005ami}, ICSI~\cite{janin2003icsi}, ISL (LDC2004S05), NIST (LDC2004S09) and SPINE1\&2 (LDC2000S87, LDC2000S96, LDC2001S04, LDC2001S06, LDC2001S08). The 8k sampled telephone data covers six monolingual CallHome sets, including Arabic (LDC97S45), English (LDC97S42), German (LDC97S43), Japanese (LDC96S37), Mandarin (LDC96S34) and Spanish (LDC96S35). MUSAN~\cite{snyder2015musan} and RIRS~\cite{ko2017study} corpora are employed for data augmentation. 

For validation on the DIHARD \rom{2} dev set, we regard it as a held-out dataset. It means neither threshold tuning nor model adaptation is performed. As for evaluation on the DIHARD \rom{2} evaluation (eval) set, we take the dev set to finetune our models and improve the performance. 

\subsection{VAD}
\subsubsection{WebRTC}
WebRTC~\cite{webrtc2011} is the official VAD baseline for Track2. Raw audios are split into frames with 20ms duration. For each input frame, WebRTC generates output 1 or 0, where 1 denotes speech and 0 denotes nonspeech. An optional setting of WebRTC is the aggressive mode, an integer between 0 and 3. 0 is the least aggressive about filtering out nonspeech while 3 is the most aggressive.

\subsubsection{ResNet-LSTM VAD}
\label{sec:vad}
We propose a DNN based approach for the VAD task. The network structure, shown in Figure~\ref{fig:vad}, consists of a ResNet module, multiple Bi-LSTM layers and linear layers. Our motivation is that the ResNet module generates representative feature mappings for speech and nonspeech, and the Bi-LSTMs capture sequential information. The input is a long sequence of frame-wise features. Every 8 frames in the sequence are packed and fed into the ResNet, generating multi-channel feature maps $\boldsymbol{F}\in \mathbb{R}^{C\times H \times W}$. We apply the global average pooling (GAP) on each channel and get a $C$-dimensional vector. Next are Bi-LSTM layers to capture forward and backward sequence information. Finally, outputs from the Bi-LSTMs pass through linear layers connected with the sigmoid function, and generate the speech likelihood. 

To be specific, we extract 32-dimensional log-mel-filterbank (fbank) features from audios by 25ms length and 10ms step. An input sequence contains $200\times 8=1600$ frames. ResNet18 is employed and its channel width is set to \{16, 32, 64, 128\}. Two Bi-LSTM layers are stacked, each with 64 units per direction and a dropout rate of 0.5. Followings are two linear layers with 64 units and 1 unit, respectively. Details about model parameters and output size are shown in Table~\ref{tab:vad_para}. Training sets include 16k meeting data and 8k telephone data with augmentation, as described in Section~\ref{sec:data}. They are all downsampled to 8k sample rate. The stochastic gradient descent (SGD) optimizer and the binary cross entropy (BCE) loss are employed. The learning rate is initialized as 0.1 and decreases by a factor of 1/10 every 20 epochs. The training process terminates after 60 epochs. For evaluation, we take the dev set to finetune the model for 10 more epochs.

\newcommand{\blockb}[3]{\multirow{2}{*}{\(\left[\begin{array}{c}\text{conv 3$\times$3, #2}\\[-.1em] \text{conv 3$\times$3, #1}\end{array}\right]\)$\times$#3}}
\newcommand{\blocks}[3]{\multirow{2}{*}{\(\left[\begin{array}{c}\text{conv 3$\times$3, #2}\\[-.1em] \text{conv 3$\times$3, #1}\end{array}\right]\)$\times$#3, /2}}
\begin{table}[!htb]
 \centering
 \caption{Parameters and output size of the ResNet-LSTM VAD.}
 \label{tab:vad_para}
 \vspace{2mm}
 \begin{tabular}{|c|c|c|}
 \hline
 Layer & Structure & Output size \\ \hline
 Inputs & - & $8\!\times\!32$ \\ \hline
 \multirow{9}{*}{ResNet} & conv $3\!\times\!3, 16$ & $16\!\times8\!\times\!32$ \\ \cline{2-3}
 & \blockb{16}{16}{2} & \multirow{2}{*}{$16\!\times\!8\!\times\!32$} \\
 & & \\ \cline{2-3}
 & \blocks{32}{32}{2} & \multirow{2}{*}{$32\!\times\!4\!\times\!16$} \\
 & & \\ \cline{2-3}
 & \blocks{64}{64}{2} & \multirow{2}{*}{$64\!\times\!2\!\times\!8$} \\
 & & \\ \cline{2-3}
 & \blocks{128}{128}{2} & \multirow{2}{*}{$128 \!\times\!1\!\times\!4$}\\
 & & \\ \hline
 Pooling & GAP & 128 \\ \hline
 \multirow{2}{*}{Bi-LSTM} & 64 units per direction, & \multirow{2}{*}{128} \\
 & 2 layers, drop=0.5 & \\ \hline
 Linear1 & $128\times 64$ & 64 \\ \hline 
 Linear2 & $64\times 1$ & 1 \\ \hline
 \end{tabular}
\end{table}

\subsection{Segmentation}
We use the uniform segmentation with overlap rather than SCD to reduce computation. In condition that the sliding window of uniform segmentation is short enough, we can assume large portions of the split segments are speaker-homogeneous. It is also mentioned in~\cite{tranter2006overview} that uniform segmentation does not significantly degrade the overall diarization performance in comparison with SCD based segmentation. In our experiments, the sliding window is 1.5s long with 0.75s step. The ground-truth speaker of a segment is the one who talks most in the central 0.75s region. 

\subsection{Speaker embedding training}
\subsubsection{i-vector}
I-vector is an unsupervised speaker embedding with generative models~\cite{shum2010unsupervised}. Essential components for an i-vector system include the universal background model (UBM) and the total variability space $\boldsymbol{T}$. For a pre-trained system, we extract features from input audios and project on UBM to calculate the statistics supervectors. Then supervectors are mapped into the low-rank subspace $\boldsymbol{T}$ as i-vectors.

We follow the dihard2018/v1 recipe in kaldi~\cite{povey2011kaldi} to train the i-vector extractor. 24-dimensional MFCCs are extracted from voxceleb1\&2 with 25ms length and 10ms step. Cepstral mean normalization (CMN) is applied. The UBM includes 2048 components, and the dimension of i-vector is 400. 

\subsubsection{x-vector}
x-vector is the DNN based speaker embedding~\cite{snyder2018x}. In the training process, frame-wise features are fed into the time-delay neural network (TDNN) for supervised learning. Outputs from TDNN are pooled over the temporal dimension and transformed into a segment-level embedding, followed by multiple linear layers. For testing cases, embeddings from the second linear layer are taken as the x-vector.

Similarly, we follow the dihard2018/v2 kaldi recipe~\cite{sell2018diarization} to train our x-vector extractor. Voxceleb1\&2 are augmented by MUSAN and RIRS corpora, and MFCCs are extracted with CMN. The dimension of x-vector is 512.

\subsubsection{Deep ResNet based speaker embedding}
The Deep ResNet structure~\cite{weicheng2019tf} is an improved version of~\cite{cai2018exploring, cai2018analysis}. It consists of three main components: a ResNet module, two parallelized pooling layers and two linear layers. First, the ResNet module transforms $L$-frame fbanks to feature maps $\boldsymbol{F}\in \mathbb{R}^{C\times H \times W}$. For each channel, the GAP layer and the global standard deviation pooling (GSP) layer are applied respectively, and outputs are concatenated together as a $2C$-dimensional vector $\boldsymbol{v}$:
\begin{equation}
 v_{1c} = \text{GAP}(\boldsymbol{F}_c) = \frac{1}{H\times W}\sum_{h=1}^H\sum_{w=1}^WF_{c,h,w}, 
\end{equation}
\begin{equation}
 v_{2c} = \text{GSP}(\boldsymbol{F}_c) = \frac{1}{H\times W}\sqrt{\sum_{h=1}^H\sum_{w=1}^W(F_{c,h,w} - v_{1c})^2}, 
\end{equation}
\begin{equation}
 \boldsymbol{v} = [v_{11}, \cdots, v_{1C}, v_{21}, \cdots, v_{1C}].
\end{equation}
$\boldsymbol{v}$ is later fed into two linear layers and the softmax function sequentially, generating speaker likelihoods. Size of the last layer equals the number of speakers in training data.

In our experiments, the training sets include voxceleb1\&2 and their augmentation, where there are 7323 speakers in total. We extract 64-dimensional fbank features from single-speaker utterances, with the number of frames $L$ ranging from 200 to 400 dynamically. ResNet34 is employed and its channel width is set to \{32, 64, 128, 256\}. The two linear layers after pooling are 128-dimensional and 7323-dimensional, respectively. We list details of model parameters and output size in Table~\ref{tab:sv_para}. The SGD optimizer and cross entropy loss are employed. The initial learning rate is 0.1, and reduces by a factor of 1/10 at the 25th and 40th epoch. Training process stops after 50 epochs. For testing cases, outputs from the first linear layer are taken as speaker embeddings.
\begin{table}[!thb]
 \centering
 \caption{Parameters of the Deep ResNet based speaker verification system.}
 \label{tab:sv_para}
 \vspace{2mm}
 \begin{tabular}{|c|c|c|}
 \hline
 Layer & Structure & Output size \\ \hline
 Inputs & - & $L\!\times\!64$ \\ \hline
 \multirow{9}{*}{ResNet} & conv $3\!\times\!3, 32$ & $32\!\times\!L\!\times\!64$ \\ \cline{2-3}
 & \blockb{32}{32}{3} & \multirow{2}{*}{$32\!\times\!L\!\times\!64$} \\
 & & \\ \cline{2-3}
 & \blocks{64}{64}{4} & \multirow{2}{*}{$64\!\times\!\frac{L}{2}\!\times\!32$} \\
 & & \\ \cline{2-3}
 & \blocks{128}{128}{6} & \multirow{2}{*}{$128\!\times\!\frac{L}{4}\!\times\!16$} \\
 & & \\ \cline{2-3}
 & \blocks{256}{256}{3} & \multirow{2}{*}{$256\!\times\!\frac{L}{8}\!\times\!8$} \\
 & & \\ \hline
 Pooling & GAP + GSP & 512 \\ \hline
 Linear1 & $512\times 128$, drop=0.5& 128 \\ \hline 
 Linear2 & $128\times 7323$ & 7323 \\ \hline
 \end{tabular}
\end{table}

\subsection{Similarity measurement}
Given speaker embeddings $\boldsymbol{x}_1, \boldsymbol{x}_2, \cdots, \boldsymbol{x}_n$ from the same audio, we compute similarity scores $S_{i, j}$ between any two embeddings $\boldsymbol{x}_i$ and $\boldsymbol{x}_j$, and construct similarity matrix $\boldsymbol{S}\in \mathbb{R}^{n\times n}$ for the clustering step.

% \subsubsection{Cosine similarity}
% The simplest way to measure score $S_{i,j}$ between $\boldsymbol{x}_i$ and $\boldsymbol{x}_j$ is to compute cosine value of their angle. In our experiments, this method is followed by spectral clustering which only accepts non-negative scores, so we normalize scores to [0, 1]:
% \begin{equation}
% S_{i,j} = \frac{1}{2}(\frac{\boldsymbol{x}_i^\top\boldsymbol{x}_j}{\|\boldsymbol{x}_i\|\cdot \|\boldsymbol{x}_j\|} + 1).
% \end{equation}
% In case that trainging set and testing set mismatch, PLDA performs better by removing channel information. 

\subsubsection{PLDA}
The PLDA score between $\boldsymbol{x}_i$ and $\boldsymbol{x}_j$ can be expressed as:
\begin{equation}
 S_{i,j} = log(\frac{p(\boldsymbol{x}_i, \boldsymbol{x}_j | \mathcal{H}_0)}{p(\boldsymbol{x}_i | \mathcal{H}_1)\cdot p(\boldsymbol{x}_j | \mathcal{H}_1)}).
\end{equation}
$\mathcal{H}_0$ hypothesizes $\boldsymbol{x}_i$ and $\boldsymbol{x}_j$ from the the same speaker while $\mathcal{H}_1$ hypothesizes $\boldsymbol{x}_i$ and $\boldsymbol{x}_j$ from different speakers. In our experiments, PLDA is trained by speaker embeddings from voxceleb1\&2 without augmentation, and whitened by the dev set.

\subsubsection{LSTM}
Noticing that PLDA handles speaker embeddings in a pair-wise and independent manner which ignores sequential information, we proposed LSTM-based scoring model in~\cite{lin2019lstm} to capture the forward and backward messages. In comparison with PLDA, scores are calculated between vector and sequence rather than vector and vector. Given speaker embeddings $\boldsymbol{x}_i$ and $\boldsymbol{x}_1, \boldsymbol{x}_{2},\cdots, \boldsymbol{x}_{n}$, we repeatedly concatenate $\boldsymbol{x}_i$ with each embedding in the sequence and feed the concatenated sequence into the LSTM network. Outputs are scores of input concatenated vectors:
\begin{equation}
 [S_{i,1}, S_{i,2}, ... S_{i,n}] = f_{\text{LSTM}}\left(
 \begin{bmatrix} \boldsymbol{x}_i\\ \boldsymbol{x}_1 \end{bmatrix},
 \begin{bmatrix} \boldsymbol{x}_i\\ \boldsymbol{x}_{2} \end{bmatrix},\cdots
 \begin{bmatrix} \boldsymbol{x}_i\\ \boldsymbol{x}_{n} \end{bmatrix}\right).
\end{equation}
For large score matrix $\boldsymbol{S}$ where $n\ge 400$, we partition it into small blocks and then perform LSTM scoring respectively, as is described in~\cite{lin2019lstm}. 

The architecture of LSTM network includes two Bi-LSTM layers followed by two linear layers. Both Bi-LSTM layers have 512 units (256 units per direction), and a dropout rate of 0.5 is added. The first linear layer is 64-dimensional and the second layer is 1-dimensional, connected with the sigmoid function to generate a similarity score between 0 and 1. Training features are speaker embeddings extracted from 16k meeting data described in Section~\ref{sec:data}. The SGD optimizer and the BCE loss are employed. The learning rate is initialized as 0.01 and decays by a factor of 1/10 every 40 epochs. The training process terminates after 100 epochs and the model is validated on the dev set. For evaluation, we further take the dev set to finetune our model for 30 more epochs. 

\subsection{Clustering}
\subsubsection{Agglomerative hierarchical clustering}
Agglomerative hierarchical clustering (AHC) is performed as the binary-tree building process~\cite{gowda1978agglomerative}. Segments are initialized as singleton clusters, and clusters with the highest pairwise similarity score are merged. The process iterates until scores between any two clusters is below a given threshold. For evaluation, we tune the threshold on the dev set to achieve the best performance. 

\subsubsection{Spectral clustering}
Spectral clustering (SC) is a graph-based clustering algorithm~\cite{von2007tutorial}. Given the similarity matrix $\boldsymbol{S}$, it considers $S_{i,j}$ as the weight of the edge between nodes $i$ and $j$ in an undirected graph. By removing weak edges with small weights, spectral clustering divides the original graph into subgraphs. As described in~\cite{von2007tutorial}, spectral clustering consists of the following steps:
\begin{enumerate}[a)]
\item Construct $\boldsymbol{S}$ and set diagonal elements to 0.
\item Compute Laplacian matrix $\boldsymbol{L}$ and perform normalization:
\begin{equation}
 \boldsymbol{L} = \boldsymbol{D} - \boldsymbol{S}, 
\end{equation}
\begin{equation}
 \boldsymbol{L}_{\text{norm}} = \boldsymbol{D}^{-\frac{1}{2}}\boldsymbol{L}\boldsymbol{D}^{-\frac{1}{2}}, 
\end{equation}
where $\boldsymbol{D}$ is a diagonal matrix and $D_i = \sum_{j=1}^{n}S_{i,j}$.
\item Compute eigenvalues and eigenvectors of $\boldsymbol{L}_{\text{norm}}$.
\item Compute the number of clusters $k$. One property of $\boldsymbol{L}_{\text{norm}}$ indicates that the number of clusters in the graph equals algebraic multiplicity of the 0 eigenvalue. In our implementation, we count the number of eigenvalues below the a given threshold as $k$. The threshold is tuned by the dev set. 
\item Take the $k$ smallest eigenvalues $\lambda_{1},\lambda_{2},...\lambda_{k}$ and corresponding eigenvectors $\boldsymbol{p}_1, \boldsymbol{p}_2,...\boldsymbol{p}_k$ of $\boldsymbol{L}_{\text{norm}}$ to construct matrix $\boldsymbol{P}\in \mathbb{R}^{n\times k}$ using $\boldsymbol{p}_1,\boldsymbol{p}_2,...\boldsymbol{p}_k$ as columns.
\item Cluster row vectors $\boldsymbol{y}_1, \boldsymbol{y}_2,...\boldsymbol{y}_n$ of $\boldsymbol{P}$ to $k$ classes by the K-means algorithm. $\boldsymbol{y}_i\in$ class $j$ indicates that segment $i$ belongs to speaker $j$. 
\end{enumerate}

\subsection{Resegmentation}
Since results from clustering modules are segment-wise while the metric of DER computes error in a frame-wise manner, it unavoidably imposes loss of precision. To deal with the problem, resegmentation modules estimate distributions of clustering results and refine them by frames. 

\subsubsection{GMM resegmentation}
Given diarization outputs, the Gaussian Mixture Model (GMM) resegmentation constructs speaker-specific GMMs for each speaker by their speech. Then for each frame in the audio, we compute its posterior probabilities of belonging to different GMMs and reassign it to the GMM with the highest probability. The process iterates until converge. 

We extract 24-dimensional MFCCs from raw audios to construct speaker-specific GMMs, each with 8 components. The resegmentation process iterates for a maximum number of 5 turns.

\subsubsection{VB diarization}
The VB diarization method~\cite{diez2018speaker} constructs a HMM with eigenvoice priors. In comparison with GMM resegmentation, it takes speaker transition probablities into consideration and avoids frequent speaker turns. Besides, a UBM is pretrained and speaker-specific GMMs are adapted on the UBM. Speaker-specific GMMs share the same component weights $w_c^{ubm}$ and convariance matrices $\boldsymbol{\Sigma}_c^{ubm}$, and thus only mean vectors require adaptation. Let $\boldsymbol{u}_s$ be the mean supervector concatenated by GMM component means for speaker $s$, and $\boldsymbol{u}_{ubm}$ be the mean supervector of UBM. The algorithm assumes
\begin{equation}
 \boldsymbol{\mu}_s = \boldsymbol{\mu}_{ubm} + \boldsymbol{T}\boldsymbol{z}, 
\label{eq:vb}
\end{equation}
where $\boldsymbol{z}$ is the low-dimensional latent vector and subject to normal distribution $\mathcal{N}(\boldsymbol{0}, \boldsymbol{I})$. $\boldsymbol{T}$ is the total variability subspace pretrained from i-vector systems. From e.q.~\ref{eq:vb} we can infer the distribution of $\boldsymbol{\mu}_s$:
\begin{equation}
 \boldsymbol{\mu}_s \sim \mathcal{N}(\boldsymbol{\mu}_{ubm}, \boldsymbol{T}\boldsymbol{T}^\top).
\end{equation}
It further imposes priors for speaker-specific GMMs. 

In our experiments, we employ the tool released by~\cite{diez2018speaker}. UBM with 1024 components are trained on 24-dimensional MFCCs and $\boldsymbol{z}$ is 400-dimensional. Other parameters for the VB algorithm are: maxIters=1, downsample=3, loopProb=0.99, statScale=0.3.

\subsection{Speech overlap detection}
In Section 2, we point out high overlapped errors for DER metric through metadata analysis. To cope with the problem, we attempt to carry out overlap detection experiments. The model structure, data and training configurations are all the same as those in ResNet-LSTM VAD. The only difference is that we remove the nonspeech regions and mark overlapped speech and non-overlapped speech as 1 and 0, respectively. For testing cases, when a segment is discriminated as overlapped speech in audios, we extend its boundary by $\pm$20 frames and take all speakers appearing in the extended segment as labels of the original segment. 

\section{Results and discussions}
\subsection{VAD}
To have a direct view at VAD performance, we carry out independent evaluation on WebRTC and our ResNet-LSTM based VAD. Ground-truth VAD labels of the eval set are available in Track1 but should be held-out in Track2, so the experiments are performed after the competition. Since the VAD model works as a submodule for the whole diarization system, we focus on the accuracy rate as the evaluation metric. Results are shown in Table~\ref{tab:vad_res}. 

Without adaptation, the ResNet-LSTM VAD model only gains slightly higher accuracy than the WebRTC baseline. When we take the dev set to finetune the model, performance on the eval set significantly improves to 91.4\%. It mainly results from the fact that our training data is drawn from meeting and telephone domains while the DIHARD sets covers 11 domains. Domain mismatch leads to poor performance and adaptation brings significant improvement. 

\begin{table}[!htb]
\vspace{-1mm}
\caption{Accuracy rate of different VAD modules.}
\vspace{2mm}
\label{tab:vad_res}
\centering
\begin{tabular}{cccc}
\toprule
Methods & Dev(\%) & Eval(\%)& Eval(adapt)(\%) \\ \midrule
WebRTC, mode=0 & 80.9 & 79.6 & - \\
WebRTC, mode=1 & 81.5 & 79.7 & - \\
WebRTC, mode=2 & 81.9 & 78.4 & - \\
WebRTC, mode=3 & 78.3 & 75.4 & - \\ \midrule
ResNet-LSTM & 82.2 & 80.7 & \textbf{91.4} \\ \bottomrule
\end{tabular}
\end{table}

\begin{table*}[t]
\vspace{-4mm}
 \centering
 \caption{Evaluation on different combinations of speaker embeddings, scoring, and clustering methods.}
 \vspace{2mm}
 \label{tab:res1}
 \begin{tabular}{ccccc}
 \toprule
 System ID & Speaker embedding & Scoring \& Clustering & Dev(\%) & Eval(adapt)(\%) \\ \midrule
 1 & \multirow{2}{*}{i-vector} & PLDA + AHC & 28.96 & 25.84 \\
 2 & & LSTM + SC & 25.98 & 24.72 \\ \midrule
 3 & \multirow{2}{*}{x-vector} & PLDA + AHC & 27.55 & 25.26 \\
 4 & & LSTM + SC & 23.22 & 22.03 \\ \midrule
 5 & \multirow{2}{*}{Deep ResNet}& PLDA + AHC & 23.51 & 21.7 \\
 6 & & LSTM + SC & 21.95 & \textbf{20.87} \\ \midrule
 7 & \multicolumn{2}{c}{fusion (2 + 4 + 6)} & 21.36 & \textbf{20.24} \\ \bottomrule
 \end{tabular}
 \vspace{-3mm}
\end{table*}

\subsection{Speaker embedding, similarity scoring and clustering}
In Table~\ref{tab:res1}, we compare different combinations of speaker embeddings, similarity scoring and clustering methods in Track1. It is observed that the Deep ResNet based speaker embedding outperforms i-vector and x-vector in all combinations. Besides, the backend of LSTM based scoring followed by spectral clustering behaves better in comparison with PLDA and AHC. Our best single system is System 6, which achieves a DER of 20.87\%. When we fuse System 2, 4 and 6 by averaging their score matrices, the DER further reduces to 20.24\%. 

\subsection{Resegmentation}
Resegmentation is carried out on our best single system (System 6) and the fusion system (System 7). Results are shown in Figure~\ref{tab:reseg}. In our expection, the VB algorithm should outperform the GMM method, and resegmentation modules should bring similar improvement for both systems. To our surprise, for the fusion system, diarization predictions after resegmentation does not become more accurate. The most significant improvement is gained in System 6 with VB diarization, reducing the DER by 1.65\% absolutely. The phenomenon is quite confusing, and we will keep investigating the reasons in our future works. 
\begin{table}[!htb]
\vspace{-3mm}
 \centering
 \caption{Resegmentation on System 6 and the fusion system.}
 \label{tab:reseg}
 \vspace{2mm}
 \begin{tabular}{ccc}
 \toprule
 System ID & Reseg & Eval(\%) \\ \midrule
 \multirow{3}{*}{6} & - & 20.87 \\
 & GMM & 20.52 \\
 & VB & \textbf{19.22} \\\midrule
 \multirow{3}{*}{7} & - & 20.24 \\
 & GMM & 20.18 \\
 & VB & 20.12 \\ \bottomrule
 \end{tabular}
\vspace{-3mm}
\end{table}

\subsection{Overlap detection}
The last module in our diarization system is overlap detection. Since the overlapped error is as high as 10.76\% in the dev set, it reasonable for us to assume there is around 10\% of the same type error in the eval set. Experiments are carried out on System 6 with VB diarization. Results are shown in Table~\ref{res:over}. Our attempt to detect overlapped speech only slightly reduces DERs by 0.38\% in Track1 and 0.69\% in Track2. It is still very challenging because we recalled less than 1/10 of the overlapped speech. 
% \vspace{-4mm}
% \begin{table}[!htb]
% \centering
% \caption{Overlap detection on System 6 with VB diarization.}
% \label{res:over}
% \begin{tabular}{cccc}
% \toprule
% System ID & Reseg & Overlap & Eval(adapt)(\%)\\ \midrule
% \multirow{2}{*}{6} & \multirow{2}{*}{VB} & no & 19.22 \\
% & & yes & \textbf{18.84} \\ \bottomrule
% \end{tabular}
% \end{table}
\begin{table}[!htb]
 \centering
 \caption{Overlap detection on System 6 with VB diarization.}
 \vspace{2mm}
 \label{res:over}
 \begin{tabular}{ccccc}
 \toprule
 System ID & Reseg & Overlap & Track1(\%) & Track2(\%) \\ \midrule
 \multirow{2}{*}{6} & \multirow{2}{*}{VB} & no & 19.22 & 28.59 \\
 & & yes & \textbf{18.84} & \textbf{27.90} \\ \bottomrule
 \end{tabular}
\end{table}

Our final submitted pipeline includes the ResNet-LSTM based VAD, the Deep ResNet based speaker embedding, the LSTM based scoring, spectral clustering, VB diarization and overlap detection. We achieve DERs of 18.84\% for Track1 and 27.90\% for Track2, both ranking the 2nd on the leaderboard. 

\subsection{Discussions}
\begin{figure}[!htb]
 \includegraphics[width=\linewidth]{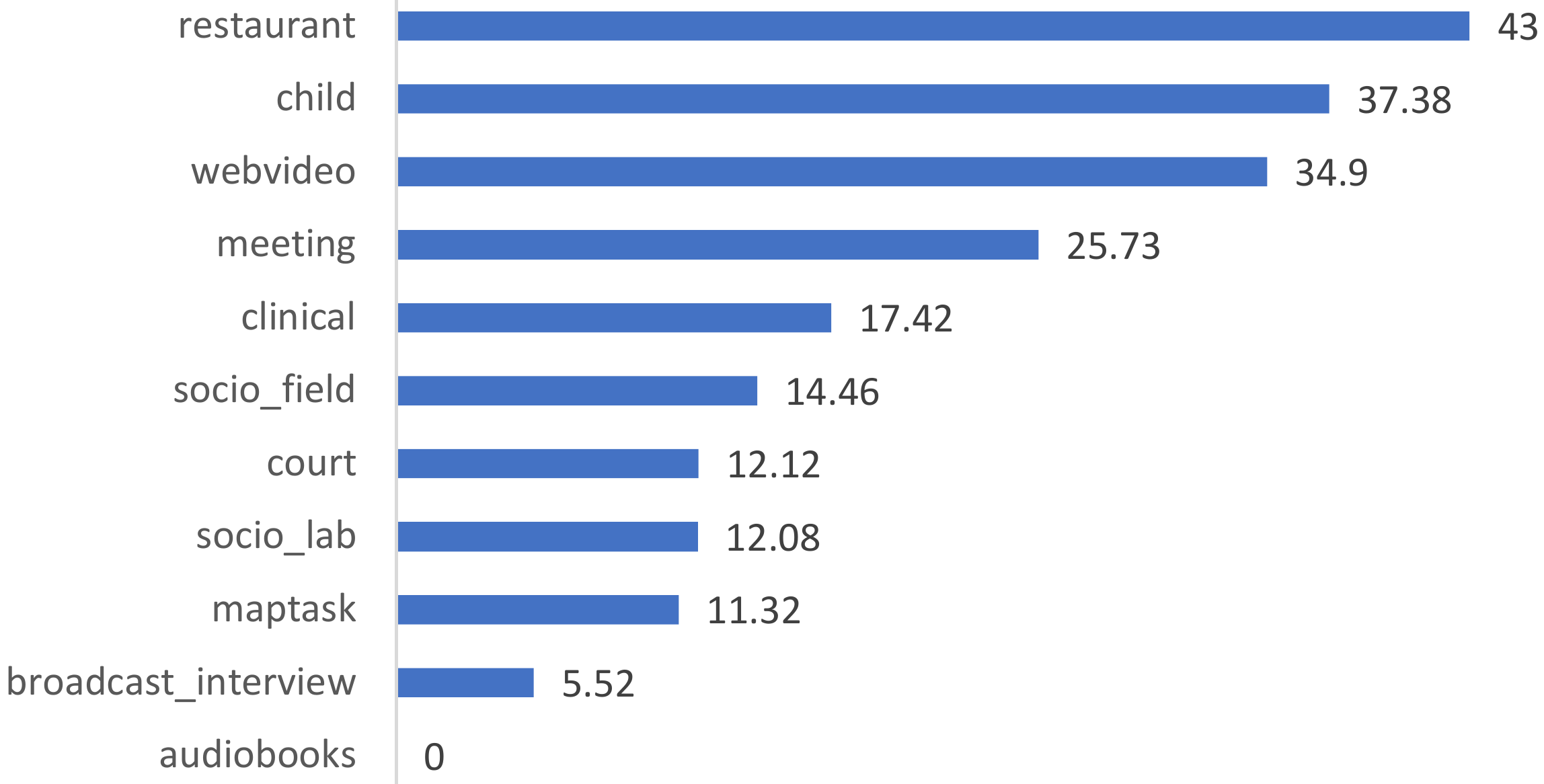}
 \caption{The performance of the proposed system in each specific domain on the DIHARD dev set.}
 \label{fig:dom}
\end{figure}

To understand how our system performs in each specific domain, we group DERs of the dev set on System 6 by domains. Results shown in Figure~\ref{fig:dom} are highly diverse between different domains. Our system performs worst in these four scenes: $restaurant$, $child$, $webvideo$ and $meeting$, three of which are discussed in Section 2 due to high overlapped errors. The $child$ domain, despite the low overlapped error, raises a high DER of 37.38\%. This is probably because the audios are drawn from Seedlings corpus where children are 6-18 months old, which is a mismatch compared with the adult speakers in our training databases. As a result, our system still performs poorly in these challenging domains. 

\section{Conclusion}
In this paper, we provide detailed introduction and compare combinations of different modules in our diarization system, and report their performance on the DIHARD \rom{2}. Our final submissions achieve DERs of 18.84\% in Track1 and 27.9\% in Track2. We further discuss DERs for specific domains, and point out that diarization is still challenging for unknown domains and domains with high overlapped errors.

\section{Acknowledgements}
This research is funded in part by the National Natural Science Foundation of China (61773413), Key Research and Development Program of Jiangsu Province (BE2019054), Six talent peaks project in Jiangsu Province (JY-074), Guangzhou Municipal People's Livelihood Science and Technology Plan (201903010040), Science and Technology Program of Kunshan City.

% \clearpage
\bibliographystyle{IEEEbib}
\bibliography{linqj}
\end{document}